\documentclass[%
superscriptaddress,
amsmath,amssymb,
 aps,prl,
 twocolumn,
 tightenlines,
]{revtex4-1}

\usepackage{graphicx,url,amssymb,amsmath,longtable,rotating,color,units,wasysym,subfigure,epsfig,hyperref,epstopdf}
\usepackage[dvipsnames]{xcolor}
\usepackage{enumitem}
\usepackage{lineno}

\usepackage{graphicx}
\usepackage{dcolumn}
\usepackage{bm}
\usepackage{url,hyperref}
\usepackage{array,multirow}
\usepackage{color} 
\usepackage{aas_macros} 
\usepackage{ifthen}
\usepackage{etoolbox}

\usepackage{siunitx} 
\definecolor{darkgreen}{rgb}{0.0, 0.5, 0.0}

\newtoggle{fullauthorlist}
\toggletrue{fullauthorlist}

\newtoggle{endauthorlist}
\toggletrue{endauthorlist}
\pdfoutput=1 

\usepackage{mathtools}
\usepackage[pass,paperwidth=8.5in,paperheight=11in]{geometry}

\usepackage{longtable}
\usepackage{graphicx}
\usepackage{amsmath,amssymb}
\usepackage{color}
\usepackage{units}
\usepackage{epstopdf}
\usepackage{hyperref}
\usepackage{multirow}
\usepackage{ifpdf}
\usepackage{siunitx}
\usepackage{float}

\newcommand{\beq}{\begin{equation}}
\newcommand{\eeq}{\end{equation}}
\newcommand{\bdm}{\begin{displaymath}}
\newcommand{\edm}{\end{displaymath}}

\definecolor{Gray}{gray}{0.9}
\definecolor{orange}{rgb}{0.9,0.5,0}

\mathchardef\mhyphen="2D
\def\code#1{\texttt{#1}}

\graphicspath{{./plots/}}

\begin{document}

\title{Noise Reduction in Gravitational-wave Data via Deep Learning}

\author{Rich Ormiston}
\affiliation{School of Physics and Astronomy, University of Minnesota, Minneapolis, Minnesota 55455, USA}

\author{Tri Nguyen}
\affiliation{LIGO Laboratory, Massachusetts Institute of Technology, 185 Albany St, 02139 Cambridge USA}

\author{Michael Coughlin}
\affiliation{School of Physics and Astronomy, University of Minnesota, Minneapolis, Minnesota 55455, USA}
\affiliation{Division of Physics, Math, and Astronomy, California Institute of Technology, Pasadena, CA 91125, USA}

\author{Rana X. Adhikari}
\affiliation{Division of Physics, Math, and Astronomy, California Institute of Technology, Pasadena, CA 91125, USA}

\author{Erik Katsavounidis}
\affiliation{LIGO Laboratory, Massachusetts Institute of Technology, 185 Albany St, 02139 Cambridge USA}

\begin{abstract}
With the advent of gravitational wave astronomy, techniques to extend the reach of gravitational wave detectors are desired. In addition to the stellar-mass black hole and neutron star mergers already detected, many more are below the surface of the noise, available for detection if the noise is reduced enough.
Our method (DeepClean) applies machine learning algorithms to gravitational wave detector data and data from on-site sensors monitoring the instrument to reduce the noise in the time-series due to instrumental artifacts and environmental contamination.
This framework is generic enough to subtract linear, non-linear, and non-stationary coupling mechanisms.
It may also provide handles in learning about the mechanisms which are not currently understood to be limiting detector sensitivities.
The robustness of the noise reduction technique in its ability to efficiently remove noise with no unintended effects on gravitational-wave signals is also addressed through software signal injection and parameter estimation of the recovered signal.
It is shown that the optimal SNR ratio of the injected signal is enhanced by $\sim 21.6\%$ and the recovered parameters are consistent with the injected set. 
We present the performance of this algorithm on linear and non-linear noise sources and discuss its impact on astrophysical searches by gravitational wave detectors.

\end{abstract}

\maketitle

\section*{Introduction}
The recent detections of gravitational waves from binary systems (see Ref.~\cite{AbEA2018b} for a summary of the first two observing runs) motivates technological and data analysis improvements to extend the reach of current gravitational wave detectors. 
The current network consists of the two Advanced LIGO (aLIGO) interferometers in the United States \citep{aligo}, the Advanced Virgo (adVirgo) interferometer in Italy \citep{avirgo}, GEO-HF in Germany \citep{LuEA2010}, the KAGRA interferometer in Japan \citep{kagra}, and eventually the LIGO-India detector in India \citep{Unn2013}.
Identification of gravitational wave events from binary systems is subject to transient and periodic noise sources in gravitational wave instruments.
Such noise sources may limit the significance of gravitational wave events and reduce them to sub-threshold level~\citep{AbEA2018b, Riles_2013}.
Ability to reduce noise in the instruments, thus enhancing their sensitivity, can enable the ability to identify additional events that would otherwise remain sub-threshold.

The ultimate sensitivity of the aLIGO detectors is dictated by the physics inherent to their design, such as shot noise of the laser light or thermal fluctuations of the mirror coatings and optic suspensions~\citep{aligo}. However, the performance during the observing runs is also influenced by environmental and technical noises which arise from factors such as earthquakes and the instrumentation or control of the interferometer respectively~\citep{AbEA2016g}. The confidence in the significance of any given signal and our ability to extract astrophysical information from it is directly impacted by the noise and sensitivity of the detectors at the time, and so there is a strong need to improve their performance by any available means.

In general, the performance of a single detector is characterized by separately considering the plethora of mechanisms by which non-astrophysical noise sources couple into the strain output of the instrument, such as the shot noise of the light incident on the output photodiodes or thermal motion of the arm cavity mirror surfaces. Once categorized into causally distinct groups, we can predict the instrument's performance from the incoherent sum of these noise mechanisms, and compare it to the observed steady state sensitivity. This is a crucial analysis when working to understand and improve the performance, as a diagnosis of what aspects or subsystems of the detector are the limiting factors. It also shows us where the observed noise exceeds the sum of the budgeted noise sources, and thereby where our understanding of the noise is incomplete.

In this analysis, we concentrate on the frequencies below 1000\,Hz since this range contains unexplained features.
Above 1000\,Hz, the detector is generally well-understood, and the features mostly corresponds to the shot noise at the output photodiode or harmonic lines that are not well witnessed by environmental sensors. 
At frequencies below 100\,Hz, there exists some amount of noise that remains unexplained.
There is also significant contribution from technical noise arising from the control of the suspended optics.
This is due in part to the control actuation necessary to keep the instrument well aligned over long periods of time and the gradual shifts in the beam spot positions. 
In addition, the low-frequency sensitivity of the detectors are especially important for detecting high mass binary mergers because these systems merge at low frequencies. 
For example, the recorded signal from GW150914 only spent about 200\,ms in the sensitive band of the instrument and were resolvable from about 35-250\,Hz~\citep{AbEA2016a}.

\section*{Background}
The aLIGO detectors employ numerous subsystems that control different aspects of the instrument and monitor its state. These are coordinated and operated in large part by a distributed digital control system which measures and records a large numbers of signals related to these subsystems, in addition to the main output which measures space-time strain. Thus, numerous signals are synchronously recorded along with the interferometer output, such as those from environmental sensors, mirror suspension actuation, and photodetectors. These auxiliary signals have the potential to witness coupling of unwanted noises into the interferometer, and are used in the commissioning of the detector to diagnose and mitigate such couplings.


\begin{figure}[t]
\includegraphics[width=0.99\linewidth]{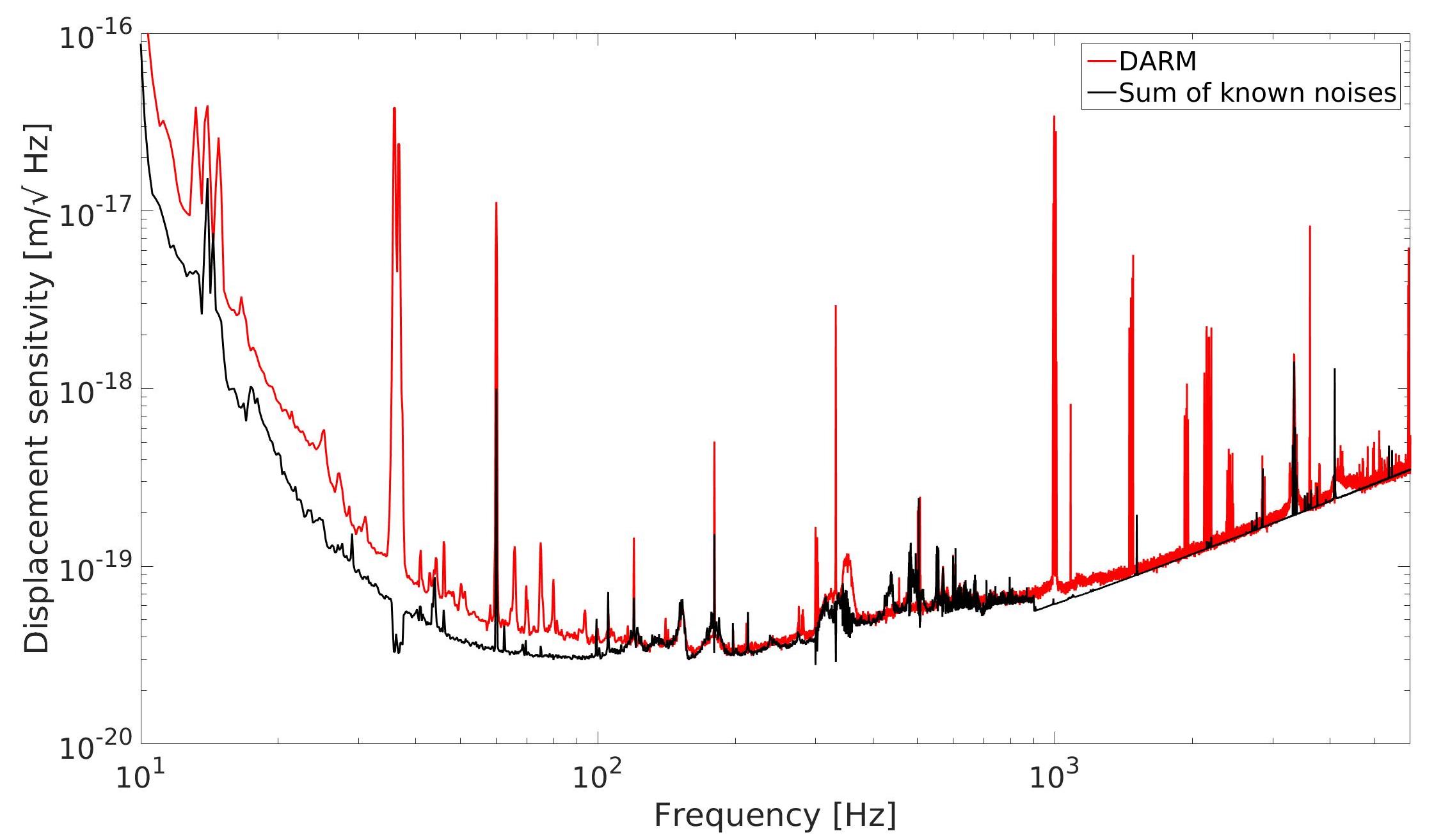}
    \caption{Estimate of the noise budget (black) and actual measured strain (red) at the Hanford interferometer for LIGO's second observing run O2. The strain of the differential arm length (DARM) was calculated using data from July 4, 2017.  \citep{Ronaldas}}. 
 \label{fig:o2_noise_full}
\end{figure}

Seismometer signals have also been used to train feed-forward subtraction filters that are run in real-time to reduce the physical motion of interferometer elements \citep{DeEA2012}. This manner of online subtraction has the strong benefit of reducing the gain or dynamic range requirements of the length and/or angular feedback systems. However, gradual changes in instrument state, such as alignment or thermal state, can cause changes in the expected couplings during an observing run, when it is preferred to make as few configuration changes to the instrument as possible. This may lead to unwanted noise making its way to the recorded strain data, despite the necessary information required to subtract it being available. Furthermore, there is the possibility of unconsidered noise couplings being present that could in principle  be predicted from other recorded signals. At this point, the only recourse is to revisit previously recorded data and attempt to regress the unwanted noise out.

One technique for reducing the noise in the strain signal post-facto using auxiliary information is Wiener filtering~\citep{Vas2001,Say2003,Davis:2018yrz,LIGOScientific:2019hgc}, a multiple-input single-output (MISO) algorithm which optimizes the mean squared difference between the subtraction target and the predicted coupled noise from multiple witnesses, taking the correlations between the witnesses themselves into account. Time domain Wiener filtering has been used successfully in terrestrial gravitational wave detectors to enhance the performance of the vibration isolation system~\citep{DeEA2012,CoMu2016} and reduce the influence of local gravitational field fluctuations~\citep{Har2015,CoHa2018}. 

Naturally, linear couplings of external disturbances into the gravitational
wave strain readout are a subset of the full dynamics of the detectors. It is therefore worthwhile to extend these regression techniques to more complicated non-linear and non-stationary couplings.  There are many known non-linear coupling mechanisms, and it is likely that more exist which have not been fully accounted for~\citep{Bose_2016}. The functional forms can vary greatly, and even modest uncertainty in the parameters involved can make it impractical to reconstruct and regress the unwanted noise.
Machine learning techniques have shown great promise at improving the sensitivity of gravitational wave data analysis. For example, Gravity Spy combines crowd-sourcing with machine learning to aid in the challenging task of categorizing all of the instrumental data transients recorded by the gravitational wave detectors \citep{ZeCo2017}. Other algorithms have risen to characterize both data transients \citep{BiBl2013,PhysRevD.95.104059} and improve gravitational wave searches \citep{GeHu2017, Tiwari_2015, Meadors_2014,mukund2020bilinear}.
This success has prompted work into developing techniques for performing linear, non-linear and non-stationary regression in the interferometer data that does not require precise a priori knowledge of all of the system parameters.

In this paper, we describe the DeepClean algorithm in which auxiliary signals are used post-facto to estimate noise couplings that existed during the recent science runs using machine learning techniques. We can use simulated gravitational wave events to characterize the performance of the algorithm and compare the performance of the noise subtraction to that of Wiener filters. We describe how the algorithms were validated to not corrupt or bias the resultant estimates of the astrophysical source parameters. Finally we detail the sensitivity improvement that results from this subtraction and the consequent improvement in the confidence in the estimates of the source parameters of select detections. 

\section*{Noise Subtraction Pipeline}
\begin{figure*}[t]
\includegraphics[width=0.9\linewidth]{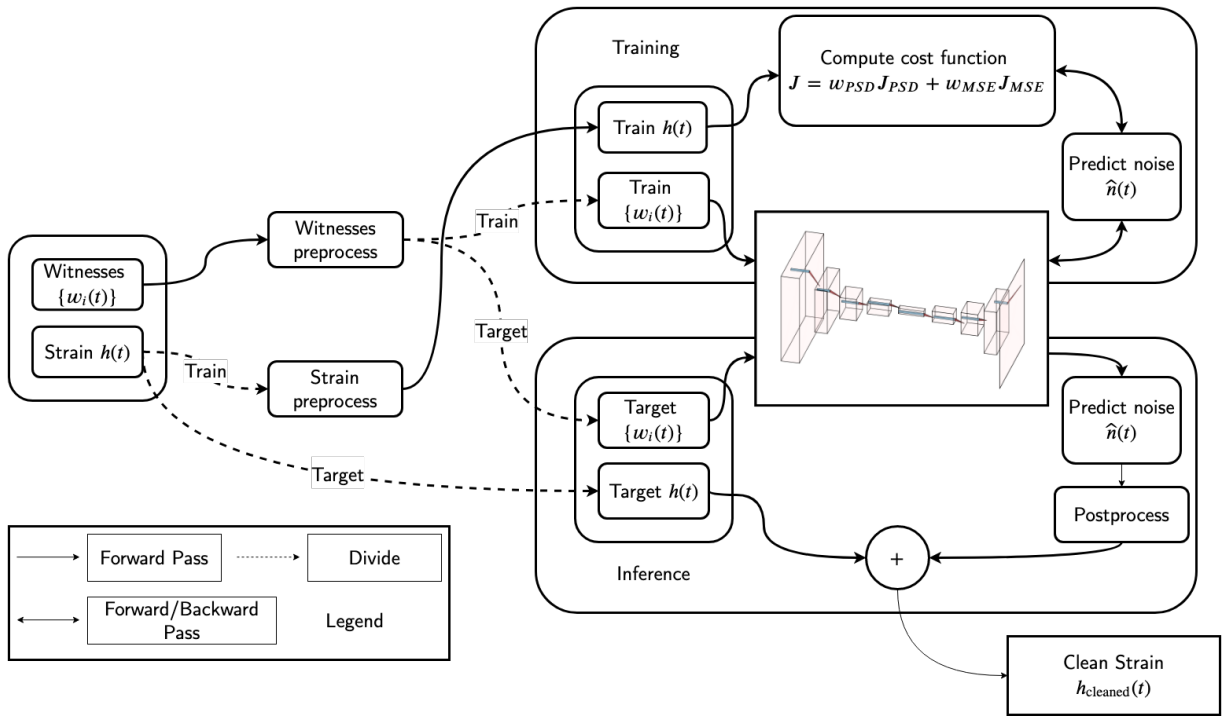}
    \caption{Workflow diagram of the noise subtraction pipeline.}
 \label{fig:deepclean_workflow}
\end{figure*}

In this section, we present the analysis pipeline applied onto data for the purpose of noise subtraction.
The method processes time-series corresponding to the gravitational wave strain data \(h(t)\) and a set of auxiliary (``witness") channels \(w_i(t)\).
The witness channels may be physical environmental monitors (PEM) or auxiliary interferometric channels that contain information about the witnessed noise and \emph{not} the astrophysical signal. 
This is a critical convenience afforded to us by our confidence that true astrophysical signals are uniquely present in the main readout signal of the interferometer; any attempted noise subtraction from a combination of witness signals may only increase or reduce the influence of noise terms, and cannot fundamentally alter any present astrophysical signals.
The algorithm employs a 1-dimensional Convolutional Neural Network (CNN) which takes in a user-specified set of witness channels at one time and subsequently outputs the predicted noise in \(h(t)\).
Input witness channels are conditioned before being fed into the CNN.
The output from the CNN (i.e., the predicted noise) is also conditioned before being subtracted from \(h(t)\).
Figure~\ref{fig:deepclean_workflow} shows the schematic of the noise subtraction pipeline.
In what follows, we will discuss the details of the algorithm implemented for this analysis.

\subsection{Formalism and Loss Function}

The gravitational-wave strain \(h(t)\) as reconstructed by the gravitational-wave detectors may be written as
\begin{equation} \label{eq:strain}
    h(t) = s(t) + n(t)
\end{equation}
where \(s(t)\) is the astrophysical signal that may be present in the data and \(n(t)\) is the noise in the detector. 
We may further subdivide the noise into 
\begin{equation} \label{eq:noise_subdivide}
    n(t) = n_w(t) + n_{nw}(t)
\end{equation}
where \(n_w(t)\) represents noise sources coupled into the witness channels \(w_i(t)\), and \(n_{nw}(t)\) represents noise sources we do not intend to subtract, which include non-removable (fundamental) noise (e.g. quantum noise, photon shot noise) and noise sources not witnessed by \(w_i(t)\).

We design the neural network to take in the witness channels and produce an estimate of the witnessed noise, which may then be filtered from the gravitational-wave strain data.
The neural network can be represented as a function \(\mathcal{F}(w_i(t);\vec{\theta})\) which maps the witness channels \(w_i(t)\) to the strain output given a set of parameters \(\vec{\theta}\).
The parameters \(\vec{\theta_i}\) are obtained by minimizing a loss function \(J\) which quantifies the difference between the predicted noise and the real witnessed noise.
Mathematically, the problem may be stated as:
\begin{equation} \label{eq:minimization}
    \vec{\theta} = \textnormal{argmin}_{\vec{\theta'}}\,J\left[h(t), \mathcal{F}(w_i(t); \vec{\theta'})\right]
\end{equation}
In our analysis, we choose the loss function to be the weighted average of the amplitude spectrum density (ASD) of the residual strain \(r(t)\) (i.e. after subtracting the predicted noise) over a frequency interval \([f_1, f_2]\).
In other words,
\begin{align} {\label{eq:psd_loss_continuous}
    J_{asd} = \frac{1}{f_2 - f_1}\int_{f_1}^{f_2} W(f) \sqrt{S[r, r](f)} df} \\
    {\label{eq:residual_strain}
    r(t) = h(t) - \mathcal{F}(w_i(t);\vec{\theta})}
\end{align}
where \(W(f)\) is a frequency dependent weighting function.
Similarly as in \citep{Vajente_2020} we choose \(W(f)\) to be the reciprocal of the ASD of the target strain $\sqrt{S[h, h](f)}$.
Since the ASD typically spans several orders of magnitude and convergence is dependent on the span of the eigenvalues of the correlation matrices of the input \citep{farhang-boroujeny_2013}, this has a whitening effect on the ASD and ensures noisy frequency bins do not dominate the loss function.
We also make a modification to set \(W(f)\) to be zero at frequencies outside the witnessed noise.
This helps the network converge faster and to a more stable solution, especially in cases where the noise source couples into the gravitational-wave strain at multiple frequencies (e.g. the 60 Hz power mains and its harmonics).
In the discrete time-series notation, the ASD loss function (Eq.~\ref{eq:psd_loss_continuous}) can be written as:
\begin{equation}\label{eq:psd_loss}
    J_{asd} = \frac{1}{M}\sum_{i=0}^{M-1} \sqrt{\frac{S[r, r][i]}{S[h, h][i]}}
\end{equation}
where \(M\) is the number of frequency bins. 

In addition to the frequency-domain loss function in Eq.~\ref{eq:psd_loss}, we find adding a time-domain loss function further improves subtraction power on spectral lines.
We choose the time-domain component of the loss function to be the mean squared error (MSE) across the time-series:
\begin{equation}\label{eq:mse_loss}
    J_{mse} = \frac{1}{N}\sum_{i=0}^{N-1} r[i]^2
\end{equation}
where \(N\) is the number of time-series samples.
Combining with the ASD loss in Eq.~\ref{eq:psd_loss}, the final loss function can be written as:
\begin{equation}\label{eq:total_loss}
    J = wJ_{asd} +  (1-w)J_{mse}
\end{equation}
where \(w\) is a weighting factor that goes from 0 to 1. 
It is a hyperparameter which can be adjusted depending on the dataset. 
In general, we found that high values of \(w\) perform better on broadband noise, and lower values of \(w\) performs better on spectral line noise. 

In practice, the MSE loss $J_{mse}$ is calculated from the full discrete time-series, while the ASD in the ASD loss $J_{asd}$ is estimated using Welch's method, i.e. averaging over the discrete Fourier Transform (DFT) of overlapping time-series segments. The number of frequency bins $M$ is chosen to be 4 to 8 times smaller than the number of time-series samples $N$ in order to obtain a non-biased estimate of the ASD. Because $M$ is proportional to the frequency resolution, we also make sure $N$ is large enough to reach a frequency resolution of at least 0.5 Hz. $M$ and $N$ remain important hyperparameters to be explored in further studies.


\subsection{Data Pre-processing}
The witness channels are pre-processed before being fed into the neural network.
Because we design the neural network to take in a time-series and output time-series of the same length, we first re-sample all witness channels, with appropriate anti-imaging filtering, to the same sampling frequency as the strain. 
By doing so, we also ensure the input and predicted time-series arrays have coincident start and stop times.
In order to save computational time, the sampling frequency of the strain is chosen such that the Nyquist frequency is just above the frequency of the witnessed noise.

Next, we apply an 8th order Butterworth bandpass filter to mitigate the power outside of the frequency band of the witnessed noise. 
For non-linear couplings, the frequency of the noise coupled into the witness channels can be different from the frequency coupled into the strain channel.
To account for this, we only bandpass the strain and not the input witnesses.

In machine learning problems, it is good practice to normalize the data to have zero mean and unit variance.
This scaling ensures that any one witness channel does not account for the majority of the error in the loss function, which would prevent the network from correctly learning the coupling of the other channels.
In addition, because the ASD of LIGO strain has a magnitude of order \(10^{-23}-10^{-21}\;\si{Hz^{-1}}\), the normalization helps prevent numerical instability when taking the ratio in Eq.~\ref{eq:psd_loss}.
We normalize both the strain channel and the witness channels.
For each channel, we compute the mean and standard deviation across the time-series.
All data samples in the time-series are then subtracted by the mean and divided by the standard deviation.
This ensures normalization does not add any unwanted features to the time-series.
The procedure is also invertible, so the network's prediction can later be easily converted back into real physical units.

To help the network learn the noise coupling more efficiently, we then divide each time-series into smaller overlapping segments.
Each training sample consists of segments from multiple witness channels. 
In machine learning literature, this step is known as data augmentation.
In our analysis, we choose a segment length to be 8 seconds.
This choice is motivated by the frequency resolution of the discrete ASD in Eq.~\ref{eq:psd_loss}.
To achieve a resolution of 0.5 Hz, we set the DFT length to 2 seconds.
Using Welch's method with a 1-second DFT overlap, we require the segment length to be at least 8 seconds.
To increase the frequency resolution, we would have to increase the DFT length, the segment length, and by extension, the computational resources of the algorithm in both time and memory.
For the noise couplings presented in the results of this paper, we find the length of 8 seconds gives the best performance in both subtraction power and run time.
We choose the overlap duration between segments to be 7.75 seconds (96.875\%) during training and 4 seconds (50\%) during inference.
Increasing the overlap duration increases the size of the dataset (and thus the computational resources) but also allows the network to make more connections and therefore characterize variants of the noise sources more effectively.
The overlap duration during inference is chosen to be smaller than during training to save computational time.

\subsection{Neural Network Architecture}\label{NNArch}
As mentioned earlier the section, the algorithm employs a 1-dimensional CNN which takes in a set of witness channels and predicts the contributed noise in the gravitational wave strain. 
The typical CNN consists of a number of convolutional layers, each employing a set of discrete window functions, or kernels, with trainable weights.
Each layer takes in an input series, which may have multiple channels.
During forward pass, each layer slides its kernel across the input and computes the dot product across all channels between the kernels and the input series within the kernel.
This locality helps the CNN learn short-term features in the data.
The output of each layer is then passed through some non-linear activation function and becomes the input of the subsequent layer.
Because the output of each layer is down-sampled, each subsequent layer sees exponentially more of the network's input.
This enables deep CNNs (with more layers) to ``remember" longer series and learn long-term features in the data.
Because the convolutional operator by definition has a built-in spatial (temporal) invariance, the CNN can also detect repeating features in the series, making it suitable to process highly periodic time-series.
In addition to convolutional layers, we employ transposed convolutional layers in our CNN.
Transposed convolutional layers \citep{dumoulin2016guide} work by exchanging the input and output of convolutional layers.
In other words, they distribute each sample of the input by the weights of the kernels and then add the resulting segments element-wise.
As a result, the output of the transposed convolution operator is up-sampled instead of being down-sampled like in the convolution operator.
The upshot is that the output of the network will be the same size as the input making the subtraction straightforward, and the transposed convolution operator adds a layer of weights which a fully connected layer or traditional up-sampling cannot.

Given the above motivations, we employ a fully-convolutional autoencoder to map the witness channels to the predicted noise.
In more detail, the input witnesses are first passed through an input convolutional layer which extracts their features and maps them to a set of output channels.
To preserve the length of the time-series, the input layer uses a stride of 1 and applies an appropriate zero padding scheme to the input witnesses.
After each layer, the series length is reduced by a factor of 2, and the number of channels increases by a factor of 2 to preserve the time complexity of the layer.
To down-sample the series, instead of using pooling layers (e.g. Max Pooling), each convolutional layer uses a stride of 2 with an appropriate zero padding scheme.
The output of the down-sampling layers is then passed through a series of transposed convolutional layers.
After each transposed layer, the series length increases by a factor 2, and the number of channels is reduced by a factor of 2.
This is done by using transposed layers with a stride of 2 and an appropriate zero padding scheme.
The output of the transposed convolutional layer is then passed through an output convolutional layer to combine all features into the predicted noise in the gravitational wave strain.
Except in the last layer, the output of each layer is normalized using Batch Normalization \citep{ioffe2015batch} and passed through an activation function before going to the subsequent layer.

The symmetrical architecture of the network is motivated by the common knowledge that each convolutional layer learns a different feature level of the input series; while the earlier layer of the CNN learns the low-level features, the deeper layer learns more advanced, high-level features. 
Therefore, we expect the first down-sampling convolutional layer to extract low-level features of the witness channels while the last transposed convolutional layer reconstructs low-level features of the predicted noise. 
Similarly, the intermediate layers will learn the intermediate-level features, and the last down-sampling convolutional layer and the first transposed convolutional layer will learn the high-level features.

In our analysis, we use three convolutional layers for down-sampling (not including the input and output convolutional layers) and three transposed convolutional layers for up-sampling. 
We set the kernel size of all layers to be 7.
The number of output channels of each layer (including the input and output layers) is 8, 8, 16, 32, 32, 16, 8, 1 respectively.
For the activation function, we use \(Tanh\).
The input and output dimension of each layer then depends only on the dimension of the input, i.e. the number of witness channels and length of the time-series, which we vary depends on the noise couplings. 
An example of the network architecture is shown in Fig~\ref{fig:60Hz_NN_diagram} in the Appendix.
We have tried different network architectures and other activations such as \(ReLU\) and \(Sigmoid\) and found that they give similar results. 

\subsection*{Training and Inference}

The pipeline can be divided into two parts: training and inference. 
During training, data are fed into the network by mini-batches wherein each batch consists of 32 training samples.
The network then computes the loss function in Eq. \ref{eq:total_loss} (averaging over the mini-batch) and its gradient with respect to the network's parameters \(\vec{\theta}\).
The parameters are updated accordingly using a first-order stochastic gradient descent algorithm.
Training terminates when gradient descent converges, which typically takes about 5-10 iterations over the training data (``epochs"), or when it reaches a set maximum number of 50 epochs.
The gradient descent algorithm used in our analysis is ADAM \citep{kingma2014adam} with default hyperparameters.
Also, we find reducing the learning rate of ADAM by a factor of 10 every 5 epochs helps the network converge to a lower loss value and improves the subtraction.
During inference, the network takes in the witness channels and produces the predicted noise.
This is processed further using the procedure described in the subsequent section before being subtracted from the gravitational wave strain.
To prevent over-fitting, which occurs when the network learns features presented uniquely in the training data without being able to generalize to a broader dataset, the training data do not overlap with the inference data.
In addition, at the end of every epoch, we compare the loss on the training and inference set and stop the gradient descent if the network is trading performance on the inference set for performance on the training set.
The duration of the training data is greatly dependent on the properties of the witnessed noise, such as the complexity of its coupling function to the gravitational-wave strain.
More complex coupling functions require longer training data.

Because noise coupling functions in LIGO are typically non-stationary, we design our network to be small such that frequent re-training does not take a significant amount of run time.
Using the architecture described above, the training time on 300-1024 seconds of training data (using 8-second segments with 7.75-second overlapping) takes about 2-6 minutes (including data pre-processing) on an NVIDIA TITAN X (Pascal) GPU.
Once trained, inference on 1024-3600 seconds of data takes a few seconds on the same GPU.
Because both the training and inference time are much less than the duration of the corresponding dataset, the algorithm can be applied for both offline and real-time subtraction.
Because the algorithm takes in raw data with no featurization, it is also easy to implement in real time.
We expect the training time to scale linearly with the duration of the training set.
The number of trainable parameters in the network also contributes significantly to the training time.
Because the input layer maps the witness channels to a fixed-dimensional subspace, increasing the number of witness channels only increases the run time of the input layer, which is a small fraction of the run time of the network.
However, whenever possible, we do not include irrelevant witness channels into the input because they hinder gradient descent convergence and possibly add uncertainty to the subtraction.


\subsection{Output Data Post-processing}

As mentioned earlier, the output of the neural network is processed before being subtracted from the original target strain.
During inference, the network takes in the witness channels in 8-second segments with a 4-second (50\%) overlap between segments and predicts the noise in the strain.
We apply a Hann window to each segment before adding them together to reduce edge-effects.

Because the network is trained on the normalized strain, the predicted noise will be in the dimensionless units.
We convert the noise back to the gravitational-wave strain unit by scaling all data samples up by the standard deviation and adding the mean computed in the normalization step during data pre-processing.

During training, the weighting function \(W(f)\) in Eq.~\ref{eq:psd_loss_continuous} is set to zero for frequencies outside the band of the witnessed noise. 
It is reasonable to assume that prediction made by the network at these frequencies will only add noise to the gravitational-wave strain.
We therefore bandpass all these frequencies using an 8th order Butterworth bandpass filter.

As mentioned earlier, to save computational time we down-sample the strain such that the Nyquist frequency is just above the frequency of the witnessed noise.
The predicted noise will have the same sampling frequency as the sampling frequency chosen during training.
Because this sampling frequency might not be optimal for detection and parameter estimation of astrophysical signals, we up-sample the noise, with an appropriate anti-imaging filter, before subtracting it from the full bandwidth strain.

\section*{Pipeline Performance on LIGO Data}

We applied our noise subtraction pipeline to data collected by the LIGO detectors during their second and third observing run (O2 and O3).
We choose multiple instances of data taking corresponding to different types of couplings present in order to study and quantify the performance of our pipeline.

\begin{figure*}[t]
\includegraphics[width=1.0\linewidth]{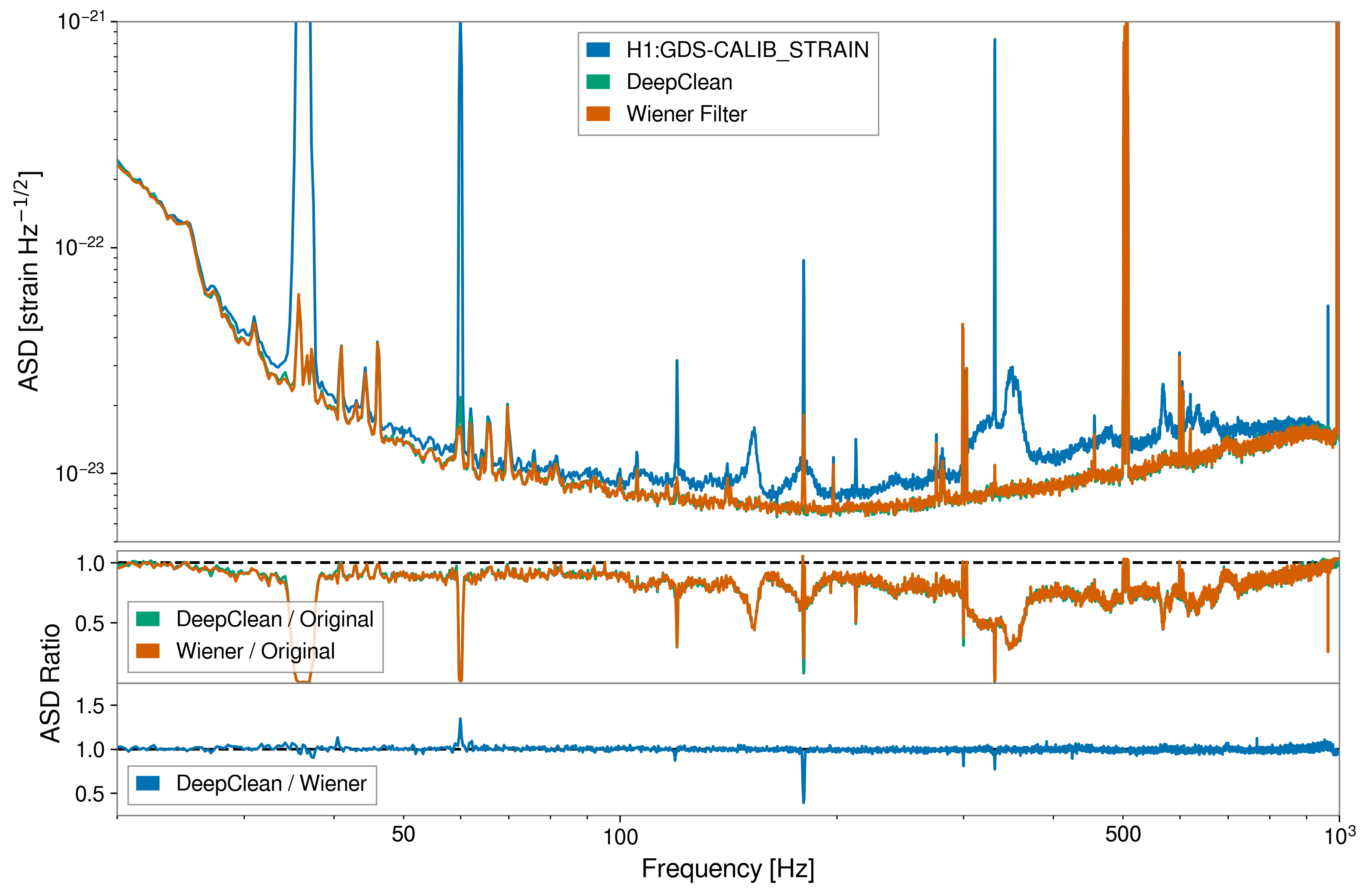}
    \caption{Comparison of the subtraction performance between a Wiener filter used to create the cleaned LIGO frames (DCH-CLEAN\_STRAIN\_C02) and the DeepClean neural network shows nearly identical results. A \(Tanh\) activation function was used which does not limit the network to simple linear connections. The results demonstrate that DeepClean can reproduce the results of linear subtraction when provided with the same channels and without overfitting the data.}
 \label{fig:wf}
\end{figure*}

\begin{figure}[t]
\includegraphics[width=1.0\linewidth]{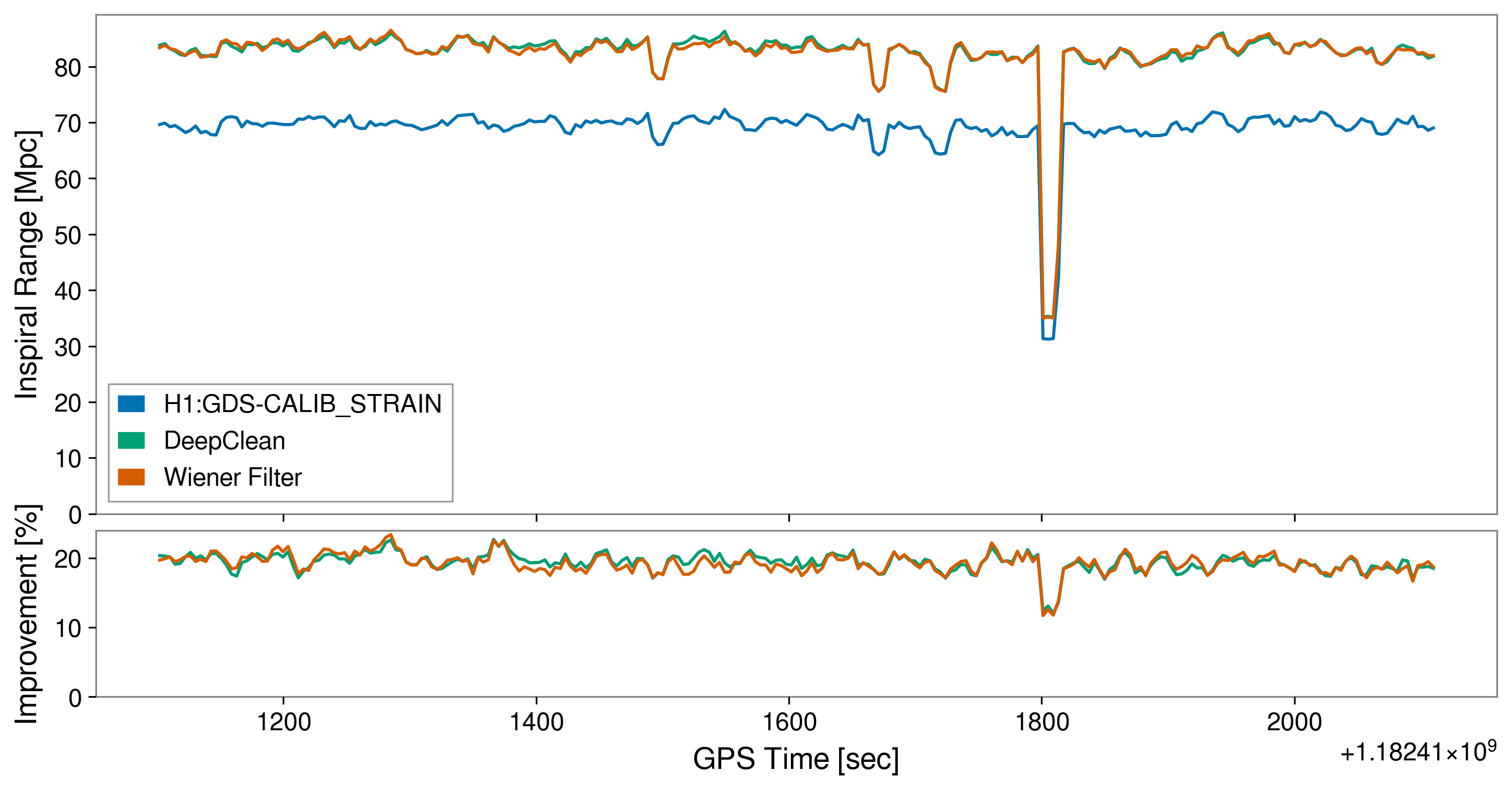}
    \caption{The increase in the binary neutron star inspiral range after the linear subtraction is seen to be $\sim 20\%$.}
 \label{fig:lho_linear}
\end{figure}

\subsection*{O2 Jitter Noise}
We first explore subtraction on the calibrated
output data from the LIGO Hanford detector during the second observing run.
This data was subsequently cleaned using a Wiener filter to linearly regress the data over LIGO's second observing run \cite{Davis:2018yrz}.
We aim to compare our method with such Wiener filter subtraction that has been adopted by the LIGO search pipelines.
The broadband linearly-coupled noise in this analysis comes from fluctuations of the pre-stabilized laser beam in size and angle \cite{Driggers_2019} and comprises the majority of the noise removed in the frequency band from \(\sim10^2-10^3\) Hz \cite{Ronaldas}.
In addition to the beam jitter noise, narrowband spectral features such as the 60 Hz power line and its harmonics, as well as the interferometer's calibration lines are also removed. 
As input to the network we use an identical set of auxiliary channels as in \cite{Davis:2018yrz}.
These channels are known to have strong linear couplings into the output strain data stream. 
However, we do not limit the network to strictly linear interactions.
As mentioned in the previous section, the network instead utilizes a non-linear \(Tanh\) activation function and therefore could in principle discern additional non-linear couplings within these channels.

We present an example of the O2 linear subtraction in Figure~\ref{fig:wf} and Figure~\ref{fig:lho_linear}.
In this example, the network is trained on the 1024 seconds of data starting at GPS time 1182410770 (2017-25-06 07:25:52 UTC) and subtracts on the 1024 seconds of data starting at GPS 1182411794 (2017-25-06 07:42:56 UTC).
The pre-processing and post-processing procedures, as well as the network architecture, are described in the previous section, with all witness and gravitational-wave strain channels re-sampled to a sampling frequency of 2048 Hz.
In addition to these steps, we remove all spectral lines (power mains and calibration lines) before subtracting the broadband beam jitter noise.
For each spectral line (or group of spectral lines, such as the calibration lines near 37 Hz), we train a different network (using the appropriate witness channels) and combine the outputs.
Because the lines are at different frequencies, they can be removed simultaneously by parallelizing the training/inference process. 
To remove the broadband beam jitter noise, a final network is trained using the interferometer strain data with the spectral lines removed.
Figure~\ref{fig:wf} shows the ASD(s) of LIGO Hanford before and after the subtraction, computed based on the inference data (and not the training data).
The improvement of the BNS inspiral range due to this linear noise subtraction is shown in Figure~\ref{fig:lho_linear}. 
The result of the Wiener filter subtraction from \cite{Davis:2018yrz} is also shown for comparison.
The range obtained from DeepClean is similar to within 1 - 2 \% of that from the Wiener filter result, suggesting that the network has learned the coefficients of the optimal MSE filter and captured physical couplings without over-fitting or adding any additional noise. 


\subsection*{O3 60 Hz Sidebands}
DeepClean is not limited to linear couplings unlike the Wiener filter.
The non-linear activation function allows the algorithm to learn arbitrarily high order couplings of the input data.
One such example of non-linear and non-stationary couplings is the modulation of a low-frequency signal from LIGO's alignment sensing and control (ASC) system with the 60 Hz line of the power mains.
This coupling produces sidebands around the central frequency.
In previous work by \cite{Vajente_2020}, the ASC system channels and a witness to the power mains have been used to subtract these sidebands during the third observing run O3.

\begin{figure}[t]
\includegraphics[width=1.0\linewidth]{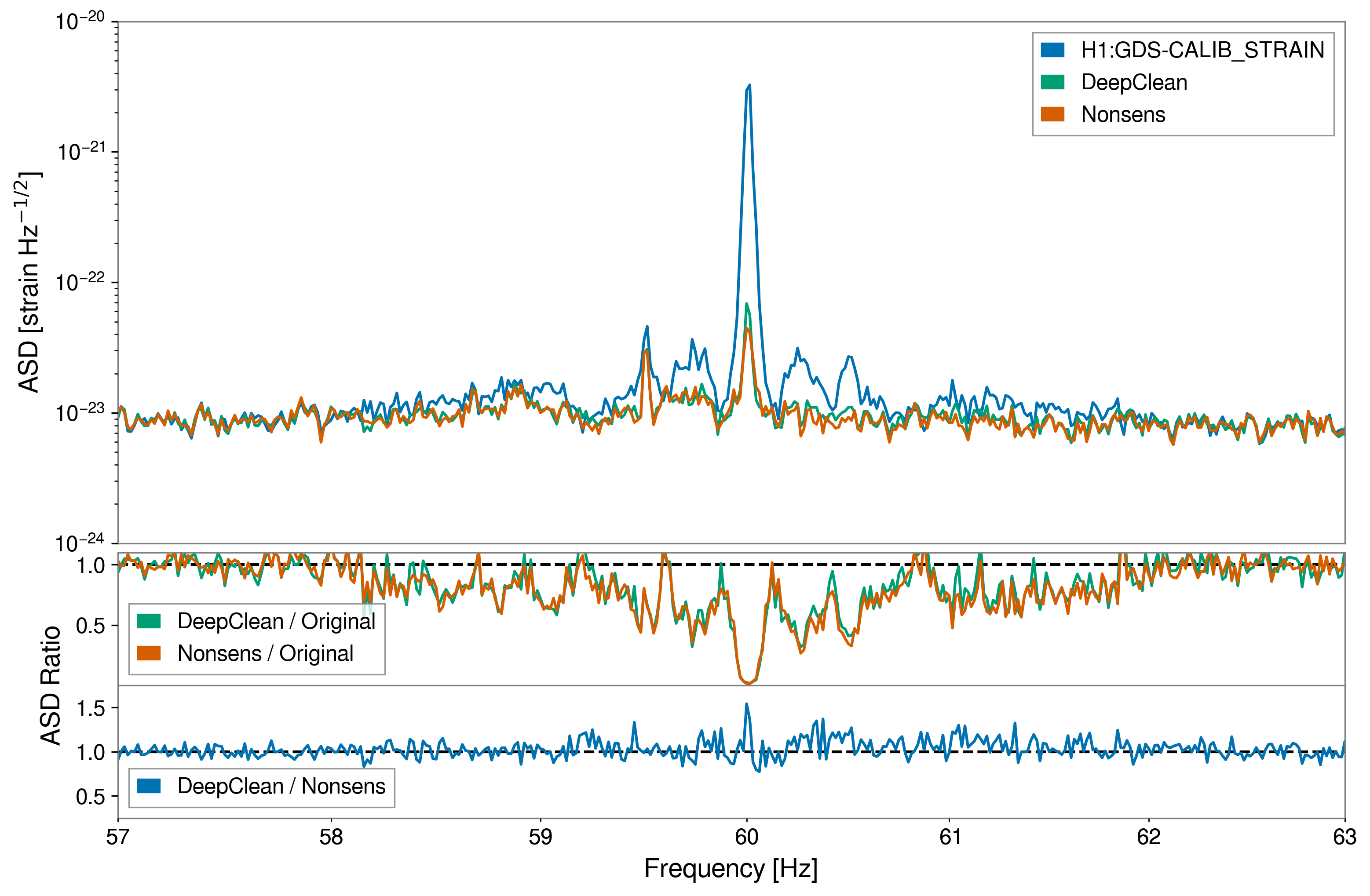}
    \caption{Since the network is not restricted to linear couplings, non-linear noise such as the modulation of the 60 Hz power line by the ASC system may be effectively and safely removed. The non-stationary subtraction in \cite{Vajente_2020} is also shown for comparison.}
 \label{fig:60Hz_before_after_asd}
\end{figure}

We have benchmarked our pipeline on the same dataset to compare the two methods.
Using the same set of witness channels as in \cite{Vajente_2020} which uses an analytic modulated linear adaptive filter in the frequency domain, we show in Figure~\ref{fig:60Hz_before_after_asd} that our network is capable of removing non-linear and non-stationary couplings such as these.
In this example, the neural network is trained on 1024 seconds of data starting at GPS time 1243926522 (2019-07-06 07:08:24 UTC) and subtracts on the 1024 seconds of data at GPS time 1243927546 (2019-07-06 07:25:28 UTC).
All channels are re-sampled to a sampling frequency of 1024 Hz.
Note that the network subtracts both the linear coupling (central peak at 60 Hz) and the non-linear and non-stationary coupling (sidebands) at the same time.

%
%
%

\section*{Parameter Estimation and Network Safety}
The procedure outlined above is carried out for each gravitational wave time-series separately, i.e. the ones from the LIGO detectors in Hanford and Livingston.
If the performance of the trained filters does not add noise and is either consistent with known results from analytic methods or subtracts spectral features of the target channel in a manner consistent with the features of the witness channels, then those filters are safe to apply to the strain data during the time of interest. The result of the filters is the production of a new strain time series which should have increased fidelity to the gravitational-wave strain signal incident on the instruments.
One way of assessing the ability of our method to denoise gravitational-wave time series is by invoking parameter estimation methods on a set of astrophysical signal waveforms that are injected via software and for signals in which the true astrophysical parameters are known {\it{a priori}}. 
In this way, we can test whether this noise subtraction scheme is legitimately reducing unwanted technical noise without distorting the measured gravitational-wave signals. We use the DeepClean algorithm to filter noise from a stretch of data which contains an astrophysical software injection. Then we check that the resultant posterior parameter estimation distributions are consistent with those from the pre-subtraction strain signal and not significantly biased away from the known injected parameters. 


\begin{figure}[htp]
\includegraphics[width=1.0\linewidth]{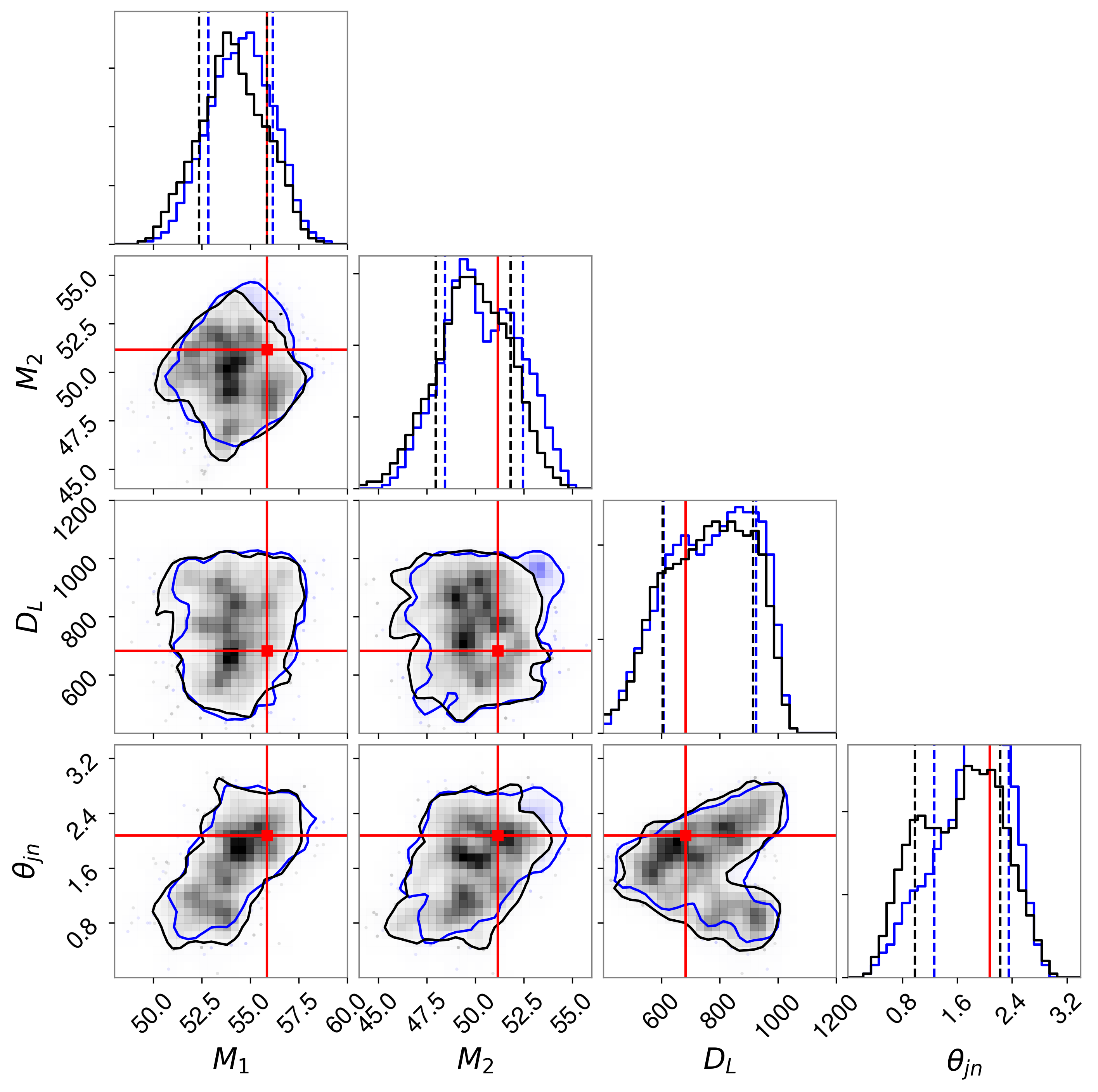}
    \caption{Corner plot showing the parameter estimation of the O2 data from the Hanford detector after cleaning the data with the DeepClean neural network and using the same auxiliary channel list as was used in the linear cleaning analyses.}
 \label{fig:PE_corner}
\end{figure}

For this study, we injected non-spinning binary black hole (BBH) signals into the gravitational-wave strain and compared the recovered source parameters from the cleaned and uncleaned time-series. 

For the O2 beam jitter dataset, we injected 10 non-spinning BBH signals.
The signals are injected at GPS time 1182411606.
Similarly as in \cite{Driggers_2019}, each BBH has component masses \(M_1\) and \(M_2\) sampled from a uniform distribution from \([28, 64] M_\odot\) with a mass ratio \(q\) constrained to \([0.125, 1.0]\). 
The sky coordinates and orientation are sampled isotropically, and the luminosity distance \(D_L\) is sampled uniformly in comoving volume from \([70, 1540]\) Mpc.
The optimal SNR of the injected signals is ranging from 1.50 to 18.7.
In the O3 60 Hz dataset, we injected 12 non-spinning, high mass BBH signals at GPS time 1243309096 and 1244006580. 
Each injection time has 6 signals, which have a mass ratio \(q\) and total mass \(M = M_1 + M_2\) combination of (0.5, 1) and (100, 150, 300) \(M_\odot\) respectively.
The high total masses are chosen such that the signals have significant power at around 60 Hz. 
Similarly to the O2 injections, the sky coordinate and orientation of the signals are sampled isotropically, with the luminosity distance sampled uniformly in comoving volume from \([70, 1540]\) Mpc.
This results in an optimal SNR range of 0.722 to 30.3 of the injected events.
The waveform model for both the O2 and O3 injections is generated from the \code{IMRPhenomPv2} waveform approximant \cite{Khan_2016,Hannam_2014,Husa_2016}.
We applied Bayesian statistics and estimated the posterior probability distribution of the source parameters using the Dynamic Nested Sampling algorithm \code{Dynesty} \cite{Speagle_2020} implemented in the \code{Bilby} library \cite{AsHu2019}.
For this study, the posterior distribution was estimated using only the gravitational-wave strain from a single detector, i.e. LIGO Hanford.
We only recovered the masses \(M_1, M_2\), inclination angle \(\theta_{jn}\), and the luminosity distance \(D_L\) of the each gravitational-wave signal.
All quantities are reconstructed within \(3\sigma\) of their true values.
In the O2 beam jitter dataset, the 90\% credible intervals of the reconstructed quantities shrink by approximately 7.25\% when comparing them with the ones obtained from the original strain.
In addition, we observed an average increase in the recovered optimal SNR of about 21.6\%.
We attribute this to the improved noise spectrum our method provides.
Figure~\ref{fig:PE_corner} shows a posterior distribution recovered from an example injection in the O2 linear dataset.
In the O3 60 Hz dataset, we did not observe any significant decrease in the 90\% credible intervals, or any substantial increase in the recovered SNR.
This is to be expected given that subtracting only the 60 Hz line and its sidebands should not significantly change the ASD. 
In all injections, the parameters recovered from the cleaned strain were consistent with the true values and those recovered from the original strain, suggesting that the network did not introduce any noise or corrupt the gravitational-wave signals.
\section*{Conclusions}
Going forward, it is evident that noise regression efforts are worth pursuing
further.
In addition to analytic methods, neural networks such as DeepClean have the extended advantage of being able to determine linear, non-linear and non-stationary couplings into the detector output without previous knowledge of the physical mechanisms of the noise.
The ability of the machine learning algorithm to successfully subtract non-linear couplings allows for network-derived filters to become a more valuable as Advanced LIGO and future detectors become increasingly sensitive to additional, more complicated noise sources and the hardware engineering limit is approached.
Given the great cost associated with the design, construction,
commissioning, and analysis of the LIGO interferometers, being able to reliably
improve the data quality through semi-automated processes will ensure a greater
science return on the investment of the scientific community and the public.
Future avenues of application could be to perform training of filters in a low-latency manner, such that a cleaned strain time series could be consistently available not long after the raw data is recorded. Running regressions in a constant online manner would also facilitate the use of cleaned data in the gravitational wave search pipelines, which require the use of the entire run's data to properly estimate the statistical significance of events over the background.
The prospects of re-running searches on previous data would be especially promising if successful non-linear regression routines are developed to capture sources such as scattering noise which was known to be a significant hindrance to the sensitivity of the aLIGO detectors during the first and third observing runs. In this case, it may be possible for marginal event candidates \cite{Abbott_2019} to be promoted to fully confident detections.
\section*{Acknowledgements}
\acknowledgements
Michael Coughlin was supported by NSF award PHY-0757058 and the David and Ellen Lee Postdoctoral Fellowship at the California Institute of Technology. Rich Ormiston was supported in part by the NSF award PHY-1806630.
\section*{References}
\bibliographystyle{unsrt}
\bibliography{references}
\appendix
\section{Neural network architecture}
In this section, we present the network architecture used in the O3 60 Hz dataset. 
For different datasets, we keep the network hyperparameters (e.g. kernel size, filter size, etc.) the same. 
Therefore, the input and output dimension of each layer depends on the dimension of the input, which depends on the number of witness channels and length of the time-series.
\begin{figure*}[t]
\includegraphics[width=0.9\linewidth]{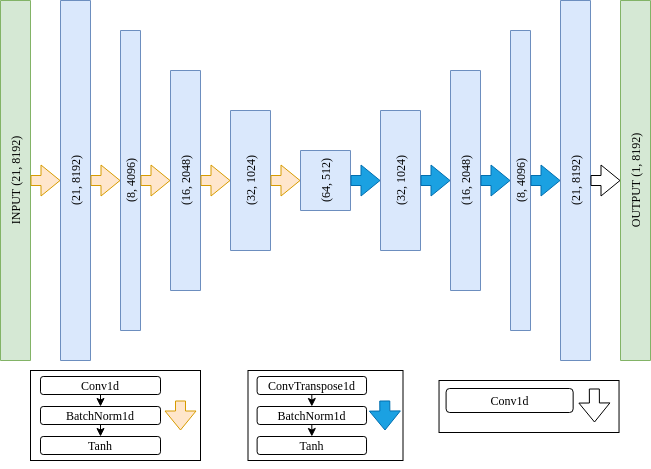}
    \caption{DeepClean architecture for the O3 60 Hz dataset. The input has 21 witness channels, including 1 PEM 60 Hz channel and 20 ASC channels. Each channel has 8192 data samples (8 seconds of data at a sample rate of 1024 Hz).}
 \label{fig:60Hz_NN_diagram}
\end{figure*}

\end{document}